\UseRawInputEncoding
\documentclass[%
 reprint,
 amsmath,amssymb,
 aps,
showkeys
]{revtex4-2}

\usepackage{graphicx}
\usepackage{dcolumn}
\usepackage{bm}
\usepackage{color}        
\usepackage{multirow}     
\usepackage[breaklinks,colorlinks,linkcolor=blue,citecolor=red,urlcolor=blue]{hyperref}          

\begin{document}

\preprint{APS/123-QED}

\title{Impact of Intrinsic Electromagnetic Structure on Nuclear Charge Radius in Relativistic 
Density functional Theory}

\author{Hui Hui Xie}
\affiliation{College of Physics, Jilin University, Changchun 130012, China}
 
 \author{Jian Li}
\email{jianli@jlu.edu.cn}
\affiliation{College of Physics, Jilin University, Changchun 130012, China}

\date{\today}

\begin{abstract}
\textcolor{black}{In this study, the effects of the nucleon's intrinsic electromagnetic (EM) structure on the nuclear charge radius \textcolor{black}{have been explored within} the framework of the relativistic Hartree-Bogoliubov (RHB) theory. It is found that the intrinsic EM structure corrections \textcolor{black}{could make an effect for \textcolor{black}{accurately} describing the evolution of nuclear charge radius.}
After taking into account the intrinsic EM structure corrections, the descriptions of the evolution of charge radii for Pb, Sn, and Cd isotopes have been improved \textcolor{black}{within} relativistic density functional theory. \textcolor{black}{Using} the Pb isotopic chain as an example, the improvement in charge radii can be primarily attributed to the intrinsic neutron and neutron spin-orbit terms of the intrinsic EM structure. \textcolor{black}{Additionally}, nuclear charge densities and corresponding isotopic evolution in Pb isotopes have been discussed.}

\end{abstract}

\maketitle


\section{Introduction}
As one of the most important nuclear properties, nuclear charge radius provides rich nuclear structure information about, \textcolor{black}{such as} nuclear shapes including deformation and shape coexistence~\cite{Bohr1975, Marsh2018}, 
pairing correlations~\cite{PhysRevLett.106.052503, PhysRevLett.110.032503, PhysRevLett.128.152501}, the presence of nuclear shell closures~\cite{PhysRevC.100.044310,Unexpectedly2016, PhysRevLett.122.192502}. \textcolor{black}{The} experimental measurement of charge radius can be derived from several electromagnetic methods that have been considerably developed over the decades~\cite{ANGELI2004185, ANGELI201369}. In particular, 
the evolutions of charge radii along the isotopic chains can be measured with high accuracy employing laser spectroscopy~\cite{CAMPBELL2016127}.

Theoretically, the nuclear density functional theory (DFT) has been widely used to provide a quite accurate global description of experimental charge radii~\cite{PhysRevC.100.044310, FAYANS200049, PhysRevC.82.035804, PhysRevC.88.011301, PhysRevC.89.054320, XIA20181,PhysRevC.104.064313}. 
For example, the Skyrme-Hartree-Fock-Bogoliubov model HFB-21 has reproduced the 782 measured charge radii with a root-mean-square (RMS) deviation of 0.027 fm~\cite{PhysRevC.82.035804}.
In comparison, it has been proposed in Ref.~\cite{PhysRevC.89.054320} that the relativistic mean field model can also reproduce well the experimental charge radii~\cite{ANGELI201369}, such as the effective interactions NL3*, DD-ME2, DD-ME$\delta$, and DD-PC1 with the RMS deviations of $0.0407$, $0.0376$, $0.0412$, and $0.0402$~fm, respectively, for all measured nuclei~\cite{ANGELI201369}.
Note that these results for charge radii $r_{\rm c}$ are conventionally calculated from point proton mean-square radii $\langle r^2\rangle_{p}$, i.e.,
\begin{equation}\label{eq-rc-p}
	r_{\rm c}=\sqrt{\langle r^2\rangle_{p}+0.64}~\mathrm{fm},
\end{equation} 
where the factor 0.64 accounts for the finite-size effect of the proton, i.e., the intrinsic proton contribution, which is a part of the intrinsic EM structure contribution. 
However, the contributions of the intrinsic neutron and the electromagnetic spin-orbit densities \textcolor{black}{were} neglected in these calculations. 

It \textcolor{black}{has been} pointed out early that the contribution of neutron spin-orbit density plays a significant role in explaining the difference of the cross sections between $^{40}$Ca and $^{48}$Ca in elastic electron scattering~\cite{BERTOZZI1972408, PhysRevC.62.054303}. \textcolor{black}{A} simple, general expression for the contribution of the spin-orbit density to the mean-square charge radius has been derived, \textcolor{black}{indicating that} it can be important in light halo-nuclei~\cite{PhysRevC.82.014320}. The significant spin-orbit contributions to both charge and weak-charge radius have been obtained from accurate calculation within relativistic mean-field models~\cite{PhysRevC.86.045503}. In addition, the corresponding contributions to the fourth-order moment of the nuclear charge density~\cite{kurasawa2019n, PhysRevC.104.024316} and charge radius in deformed nuclei~\cite{PhysRevC.103.054310} have also been discussed.
\textcolor{black}{Furthermore, recently the intrinsic EM structure contributions have been implemented within several relativistic and non-relativistic functionals for charge radii of Sn and Pb isotopes~\cite{PhysRevC.107.054307}. \textcolor{black}{However}, it is not discussed in detail how these contributions affect the charge radii. It is worthwhile to provide a detailed analysis to quantitatively demonstrate the improvement \textcolor{black}{in} describing charge radii after considering the intrinsic EM structure corrections within relativistic density functional theory. On the other hand, the pairing effect plays a crucial role in explaining the kinks (such as in Pb~\cite{PhysRevLett.74.3744, SHARMA19939,REINHARD1995467, PhysRevLett.110.032503, Bhuyan_2021,PhysRevC.91.021302, PhysRevC.105.044303} and Sn~\cite{PhysRevC.92.044307, PhysRevLett.122.192502} isotopic chains) and the odd-even staggering of charge radii in isotopic chain~\cite{PhysRevLett.110.032503, PhysRevC.104.064313, PhysRevC.91.021302}. Hence, it is also \textcolor{black}{interesting} to explore the impact of pairing effects on electromagnetic structure corrections.}

The robust performance of the relativistic DFT has been illustrated in nuclear physics by its successful description of many nuclear phenomena, see Refs.~\cite{RING1996193, VRETENAR2005101, MENG2006470, NIKSIC2011519,meng2013progress,SHEN2019103713} for a brief review. \textcolor{black}{
To provide a proper treatment of pairing correlations and mean-field potentials, 
\textcolor{black}{the relativistic Hartree-Bogoliubov (RHB) model }
has been employed in studies of many structure properties~\cite{PhysRevLett.77.3963, Meng1998NPA, VRETENAR2005101,Zhang2003632, PhysRevLett.80.460, PhysRevC.65.041302, PhysRevLett.77.3963, MENG19981, MENG2002209,PhysRevLett.77.3963,Zhang2003632, PhysRevLett.80.460, PhysRevC.65.041302, PhysRevA.107.042807, XIE2023138232, SHANG2024138527, Xie2024PhysRevC.109.034309,Geng2023,PhysRevC.110.014308}}. Recently, a global description of various ground-state properties, such as binding energies and charge radii, over the nuclear landscape, has been performed in the framework of RHB theory, and the RMS deviation of calculated charge radii from the available experimental data is $0.0358$~fm~\cite{XIA20181}. 

In this work, taking the Pb isotopic chain as an example, a detailed calculation of the intrinsic EM structure corrections, which include intrinsic nucleon contributions and the spin-orbit contributions, to nuclear charge densities and charge radii, \textcolor{black}{is performed} within the framework of the RHB theory. The impact of pairing correlations on charge radii for open-shell nuclei \textcolor{black}{as well as} the \textcolor{black}{contributions} of the intrinsic EM structure and pairing correlations \textcolor{black}{to} the evolution of charge radii will be discussed in detail. 
\textcolor{black}{To further investigate the \textcolor{black}{impact} of the intrinsic EM structure corrections, the evolution of charge radii for Pb, Sn, and Cd isotopic chains will be studied \textcolor{black}{using} several relativistic functionals.}

This paper is organized as follows. In Sec. \ref{sec-2}, the RHB method for solving the nuclear structure \textcolor{black}{is introduced}, then the nuclear charge densities \textcolor{black}{are constructed} and a more general expression of charge radii \textcolor{black}{is derived}. In Sec. \ref{sec-3}, the pairing effects on the charge density and charge radius for $^{212}$Pb are displayed. Then the \textcolor{black}{influence} of the intrinsic EM structure contributions \textcolor{black}{on} charge radii with pairing \textcolor{black}{correlations} in the Pb isotopic chain is emphasized by comparing the RHB calculations with experimental data. 
\textcolor{black}{In addition, the evolution of charge radii for Pb, Sn, and Cd isotopic chains will be compared \textcolor{black}{using} four \textcolor{black}{different} relativistic density functionals.}
Finally, a summary and outlook are presented in Sec.~\ref{sec-4}.

\section{Theoretical framework}\label{sec-2}

\subsection{Relativistic Hartree-Bogoliubov model}

Starting from the Lagrangian density, the energy density functional of the nuclear system is constructed under the mean-field and no-sea approximations. 
By minimizing the energy density functional, one obtains the Dirac equation for nucleons within the framework of relativistic mean-field theory \textcolor{black}{as} \cite{Meng2015} 
\textcolor{black}{
	\begin{equation}\label{eq-single-dirac}
		\left[\bm \alpha\cdot\bm p+V(\bm r)+\beta \textcolor{black}{\left(M+S(\bm r)\right)}\right]\psi_k(\bm r)=\varepsilon\psi_k(\bm r),
	\end{equation}
	in which $M$ is the nucleon mass; $\bm\alpha$ and $\beta$ are the traditional $4\times 4$ matrices of Dirac operators, and $\psi_k(\bm r)$ is the corresponding single-particle wave function for a nucleon in the state $k$, $S(\bm r)$ and $V(\bm r)$ are the local scalar and vector potentials, respectively. }

To describe open-shell nuclei, pairing correlations are crucial. 
The relativistic Hartree-Bogoliubov model employs the unitary Bogoliubov transformation of the single-nucleon creation and annihilation operators in constructing the quasiparticle operator and provides a unified description of both the mean-field approximation and the pairing correlations \cite{VRETENAR2005101}. 
\textcolor{black}{Following the standard procedure of Bogoliubov transformation,} the RHB equation 
\textcolor{black}{could be derived as following:} 
\begin{equation}
	\label{RHB}
	\left(\begin{matrix}
		h_D-\lambda_\tau & \Delta \\
		-\Delta^* &-h_D^*+\lambda_\tau
	\end{matrix}\right)
	\left(\begin{matrix}
		U_k\\
		V_k
	\end{matrix}\right)=E_k
	\left(\begin{matrix}
		U_k\\
		V_k
	\end{matrix}\right),
\end{equation}
where $E_k$ is the quasiparticle energy, $\lambda_\tau (\tau=n,p)$ are the chemical potentials for neutrons and protons, 
$h_D$ refers to the Dirac Hamiltonian in Eq.~(\ref{eq-single-dirac}), and $\Delta$ denotes the pairing potential, 
\textcolor{black}{which is taken as the following form,
	\begin{equation}
		\label{Delta}
		\Delta(\bm r_1,\bm r_2) = V^{\mathrm{pp}}(\bm r_1,\bm r_2)\kappa(\bm r_1,\bm r_2),
	\end{equation}
	where
	\begin{equation}
		\label{pair}
		V^{\mathrm{pp}}(\bm r_1,\bm r_2)=\frac{V_0}{2}(1-P^\sigma)\delta(\bm r_1-\bm r_2)\left(1-\frac{\rho(\bm r_1)}{\rho_{\mathrm{sat}}}\right)
	\end{equation}
	represents the density-dependent force of zero range and $\kappa(\bm r_1,\bm r_2)$ refers to the pairing tensor \cite{Ring1980}. In Eq. (\ref{pair}), $V_0$ is the interaction strength and $\rho_\mathrm{sat}$ is the saturation density of the nuclear matter.}

\textcolor{black}{With spherical symmetry imposed, the quasiparticle wave functions $U_k$ and $V_k$ in the coordinate space can be written as 
	\begin{eqnarray}
		U_k=\frac{1}{r}\left(
		\begin{array}{c}
			iG_U^{k}(r)Y_{jm}^l(\theta,\phi)\\
			F_U^{k}(r)(\pmb{\sigma\cdot \hat{r}})Y_{jm}^l(\theta,\phi)
		\end{array}\right)\chi_t(t)\nonumber\\
		V_k=\frac{1}{r}\left(
		\begin{array}{c}
			iG_V^{k}(r)Y_{jm}^l(\theta,\phi)\\
			F_V^{k}(r)(\pmb{\sigma\cdot \hat{r}})Y_{jm}^l(\theta,\phi)
		\end{array}\right)\chi_t(t).
	\end{eqnarray}
	The corresponding RHB equation can be expressed as the following radial integral-differential equations in coordinate space~\cite{Meng1998NPA}:
	\begin{eqnarray}\nonumber &
		\frac{\mathrm dG_U}{\mathrm dr}+\frac{\kappa}{r}G_U(r)-\left[E+\lambda-V(r)+S(r)\right]F_U(r)\\ \nonumber&
		+r\int r'\mathrm dr'\Delta_F(r,r')F_V(r')=0,\\ \nonumber&
		\frac{\mathrm dF_U}{\mathrm dr}-\frac{\kappa}{r}F_U(r)
		+\left[E+\lambda-V(r)-S(r)\right]G_U(r)\\ \nonumber&
		+r\int r'\mathrm dr'\Delta_G(r,r')G_V(r')=0,\\ \nonumber&
		\frac{\mathrm dG_V}{\mathrm dr}+\frac{\kappa}{r}G_V (r)+\left[E-\lambda+V(r)-S(r)\right]F_V (r)\\ \nonumber&
		+r\int r'\mathrm dr'\Delta_F(r,r')F_U(r')=0,\\ \nonumber&
		\frac{\mathrm dF_V}{\mathrm dr}-\frac{\kappa}{r}F_V (r)-\left[E-\lambda+V(r)+S(r)\right]G_V (r)\\ &
		+r\int r'\mathrm dr'\Delta_G(r,r')G_U(r')=0.\label{RHB-eq}
	\end{eqnarray}
}

After solving the relativistic Hartree-Bogoliubov equation of Eq. (\ref{RHB-eq}), the point-proton and neutron densities are obtained by summing the norm of the corresponding quasiparticle wave functions ($\tau\in\{p,n\}$),
		\textcolor{black}{
			\begin{equation}
				\label{eq:rhov}
				\rho_{\tau}(r) =\rho_{V,\tau}(r)=\sum_{k\in\tau } \frac{n_k}{4\pi r^2}\left\{\left[G_V^k(r)\right]^2+\left[F_V^k(r)\right]^2 \right\},
			\end{equation}
			where $n_k$ refers to the occupation number of the orbital $k$.
		}

		\subsection{Nuclear charge density}
		Beginning with the single-particle electromagnetic current operator~\cite{PERDRISAT2007694}
		\begin{align}
			\hat{J}_\mu=&\gamma_\mu F_{1\tau}(Q^2)+i\frac{\mu_{\tau}}{2M_\tau}F_{2\tau}(Q^2)\sigma_{\mu\nu}q^\nu\nonumber\\
			=&G_{\mathrm E\tau}(Q^2)\gamma_\mu+ F_{2\tau}(Q^2)\frac{\mu_{\tau}}{2M_\tau}\left(-\frac{Q^2}{2M_\tau}\gamma_\mu+i\sigma_{\mu\nu}q^\nu\right),
		\end{align}
		where $M_\tau$ ($\tau\in \{p,n\}$) is the mass of the nucleon, $F_{1\tau}$ and $F_{2\tau}$ are electromagnetic Dirac and Pauli form factors, respectively. $G_{\mathrm E\tau}(Q^2)\equiv F_{1\tau}(Q^2)-\mu_\tau Q^2F_{2\tau}(Q^2)/\left(4M_\tau^2\right)$ is the electric Sachs form factor, and $\mu_\tau$ is the anomalous magnetic moment ($\mu_{ n}=-1.91304273$ and $\mu_{p}=1.792847344$ for neutrons and protons, respectively~\cite{RevModPhys.93.025010}). $Q^2\equiv- q_\mu q^\mu> 0$ \textcolor{black}{with} $q_\mu$ the momentum transfer, and $\sigma_{\mu\nu}=\frac12\left(\gamma_\mu\gamma_\nu-\gamma_\nu\gamma_\mu\right)$  is the antisymmetric tensor.
		
		The nuclear charge density is uniquely related to the nuclear charge form factor \textcolor{black}{(namely the charge density distribution in the momentum space)}, which naturally includes the point-proton and neutron densities, the proton and neutron spin-orbit densities, and the single-proton and single-neutron charge densities \cite{Friar1975, PhysRevC.62.054303, PhysRevC.103.054310,kurasawa2019n}. 
		The nuclear charge form factor $F_{\rm c}$, i.e., the ground-state expectation value of the zeroth component of the electromagnetic current $\hat J_0$, can be written as 
			\begin{equation}
				F_{\rm c}(\bm q)=\int\mathrm d^3\bm r e^{i\bm q\cdot\bm r}\sum_{\tau\in\{p,n\}}\left[G_{\mathrm E\tau}(Q^2)\rho_\tau( r)+F_{2\tau}(Q^2)W_\tau( r)\right].
			\end{equation}
	The nucleon density $\rho_\tau(r)$ is obtained from Eq.~(\ref{eq:rhov}), and the spin-orbit density $W_\tau(r)$ is derived combining its general form ~\cite{PhysRevC.62.054303,kurasawa2019n} with Eq.~(\ref{RHB-eq}),
			\begin{align}
				W_\tau(r)=&\frac{\mu_\tau}{2M}\left(-\frac{\nabla^2\rho_\tau(r)}{2M}+i\nabla\cdot\langle 0|\sum_{i\in\tau}\delta(\bm r-\bm r_i)\bm\gamma_i|0\rangle \right)\nonumber\\
						=&\frac{\mu_\tau}{M_\tau}\sum_{i\in\tau}\frac{n_i}{4\pi r^2}\frac{\mathrm d}{\mathrm dr}\left\{-\frac{S(r)}{M_\tau}G_V^i(r)F_V^i(r)\right.\nonumber\\
						&\left.+\frac{\kappa_i+1}{2M_\tau r}\left[G_V^i(r)\right]^2-\frac{\kappa_i-1}{2M_\tau r}\left[F_V^i(r)\right]^2\right.\nonumber\\&+\frac{1}{2M_\tau}\left(rG_V^i(r)\int\mathrm dr'r'\Delta_F(r,r')F_U^i(r')\right.\nonumber\\&\left.\left.+rF_V^i(r)\int\mathrm dr'r'\Delta_G(r,r')G_U^i(r')\right)\right\}.
			\end{align}
			
		The nuclear charge density can be \textcolor{black}{obtained from the inverse Fourier transformation of the nuclear charge form factor} as follows:
		\begin{equation}
			\label{eq-rhoc}
			\rho_{\rm c}(r)=\sum_{\tau\in\{p,n\}}\left[\rho_{c\tau}(r)+W_{c\tau}(r)\right],
		\end{equation}
		where
		\begin{align}\label{eq-rct}
			&\rho_{c\tau}(r)=\frac1r\int_0^\infty\mathrm dx x\rho_\tau(x)\left[g_\tau(|r-x|)-g_\tau(r+x)\right],\\
			&W_{c\tau}(r)=\frac1r\int_0^\infty\mathrm dx xW_\tau(x)\left[f_{2\tau}(|r-x|)-f_{2\tau}(r+x)\right].
		\end{align}
		The functions $g_\tau(x)$ and $f_{2\tau}(x)$ are given by
		\begin{eqnarray}
			g_\tau(x)=\frac1{2\pi}\int_{-\infty}^{\infty}\mathrm dq e^{iqx}G_{\mathrm E\tau}(Q^2),\\
			f_{2\tau}(x)=\frac{1}{2\pi}\int_{-\infty}^{\infty}\mathrm dq e^{iqx}F_{2\tau}(Q^2).
		\end{eqnarray}
		Therefore, the corresponding mean-square (ms) charge radius of nuclear charge density with the intrinsic EM structure corrections is given by
	\begin{equation}\label{eq-rc}
		r_c^2\equiv\langle r^2\rangle_c=\langle r^2\rangle_p+r^2_p+\langle r^2\rangle_{\mathrm{W_{ p}}}+\frac NZ\left(r_n^2+\langle r^2\rangle_{\mathrm{W_{ n}}}\right),
	\end{equation}
    where $Z$ and $N$ represent the proton and neutron number, respectively, and $\langle r^2\rangle_{ p}$ is the point-proton mean-square radius. The ms charge radii of single proton $r^2_{ p}$ and neutron $r^2_{ n}$ account for the finite-size effect of the nucleon. 
    The rest \textcolor{black}{two} terms $\langle r^2\rangle_{\rm W_p}$ and $N/Z\langle r^2\rangle_{W_n}$ represent the proton and neutron spin-orbit contributions, respectively.  
		
		In this work, we follow Ref. \cite{PhysRevC.62.054303} and adopt the following form factors
		\begin{align}
			\label{eq-form-factor}
			&G_{\rm Ep}=\frac{1}{\left(1+r_{ p}^2Q^2/12\right)^2},\nonumber\\
			&G_{\rm En}=\frac{1}{\left(1+r_+^2Q^2/12\right)^2}-\frac{1}{\left(1+r_-^2 Q^2/12\right)^2},\nonumber\\
			&F_{ 2p}=\frac{G_{\rm Ep}}{1+Q^2/4M_{ p}^2},\nonumber\\
			&F_{\rm 2n}=\frac{G_{\rm Ep}-G_{\rm En}/\mu_{ n}}{1+Q^2/4M_{ n}^2},
		\end{align}
		with the proton charge radius $r_{ p}= 0.8414$ fm~\cite{RevModPhys.93.025010} and $r^2_\pm= r_{\mathrm{av}}^2\pm \frac12  r_{ n}^2 $, where $r_{\mathrm{av}}^2= 0.81$ fm$^2$ is the average of the squared radius for positive and negative charge distributions and $ r_{ n}^2 =-0.11$ fm$^2$ \cite{atac2021measurement} is the ms charge radius of the neutron.

\section{Results and discussion}\label{sec-3}
To investigate the effects of intrinsic electromagnetic structure on the nuclear charge radii and charge densities in the Pb isotopes, the wave functions of the nuclear ground states \textcolor{black}{are} solved from the RHB calculations \textcolor{black}{in the coordinate representation~\cite{XIA20181}}. The Box size $R_\mathrm{box}$ = 20 fm, the mesh size $\Delta r$ = $0.1$~fm, and the angular momentum cutoff $J_\mathrm{max} = 19/2 \hbar$ are used in RHB calculations. In the present systematic calculations, the relativistic density functional PC-PK1~\cite{PhysRevC.82.054319} is adopted for the particle-hole channel, and pairing strength $V_0$ \textcolor{black}{is taken} $-342.5$~MeV fm$^3$ for both neutron and proton \textcolor{black}{with a cutoff energy of $100$ MeV for the pairing window}. More numerical details can be found in Ref.~\cite{XIA20181}.

\begin{figure}
	\centering
	\includegraphics[width=8.5cm]{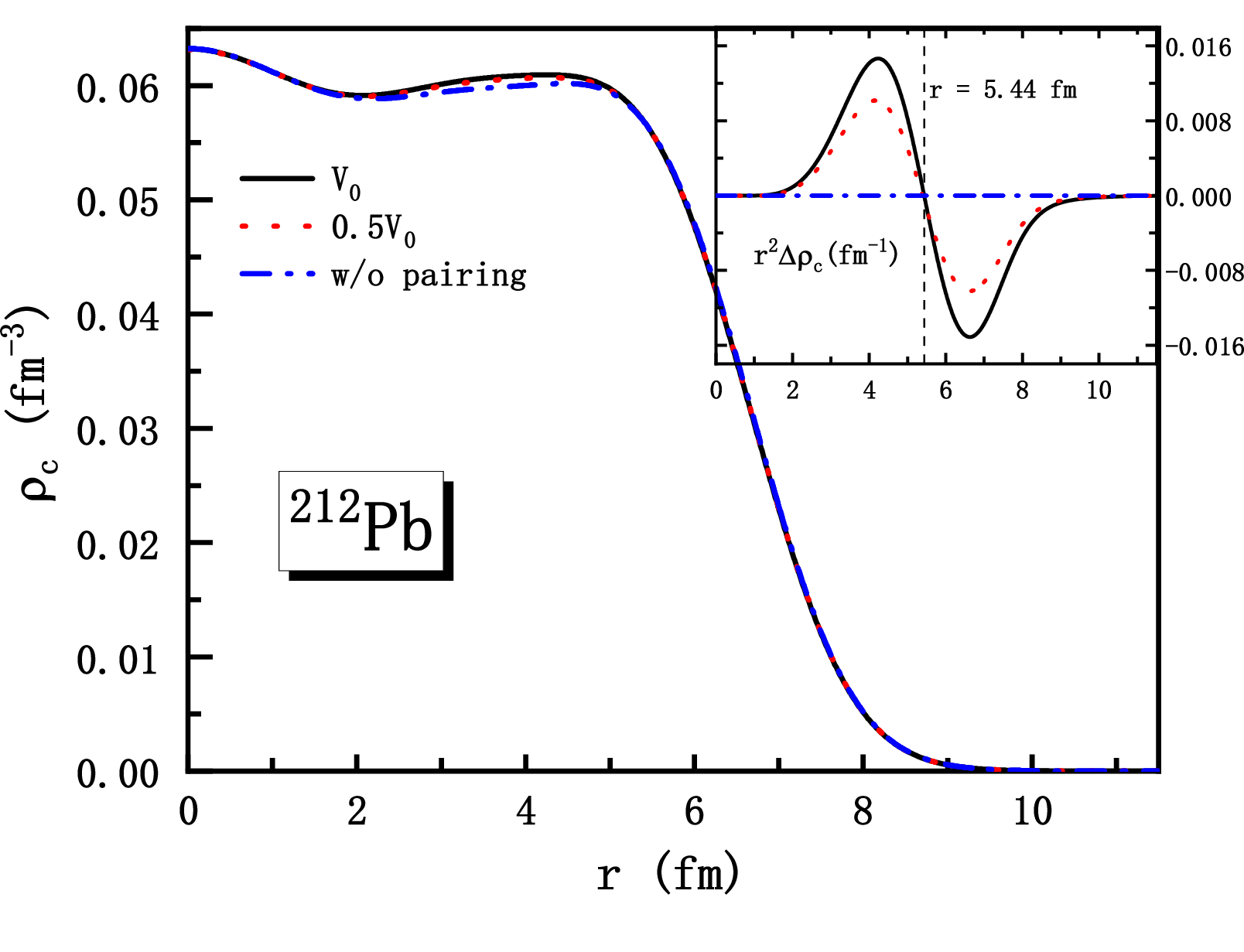}
	\caption{(Color online) Comparison of charge density for $^{212}$Pb from RHB calculations \textcolor{black}{using PC-PK1 interaction}~\cite{PhysRevC.82.054319} with pairing strength 0.5$V_0$ and $V_0$ as well as the results without pairing \textcolor{black}{correlations}. \textcolor{black}{The inset shows $r^2\Delta \rho_c$, where $\Delta \rho_c$ denote the deviation of the charge densities with pairing strength $0.5V_0$ and $V_0$ from the result without pairing \textcolor{black}{correlations}. 
 }}
	\label{fig:fig1}
\end{figure}

\begin{figure}
	\centering
	\includegraphics[width=8.5cm]{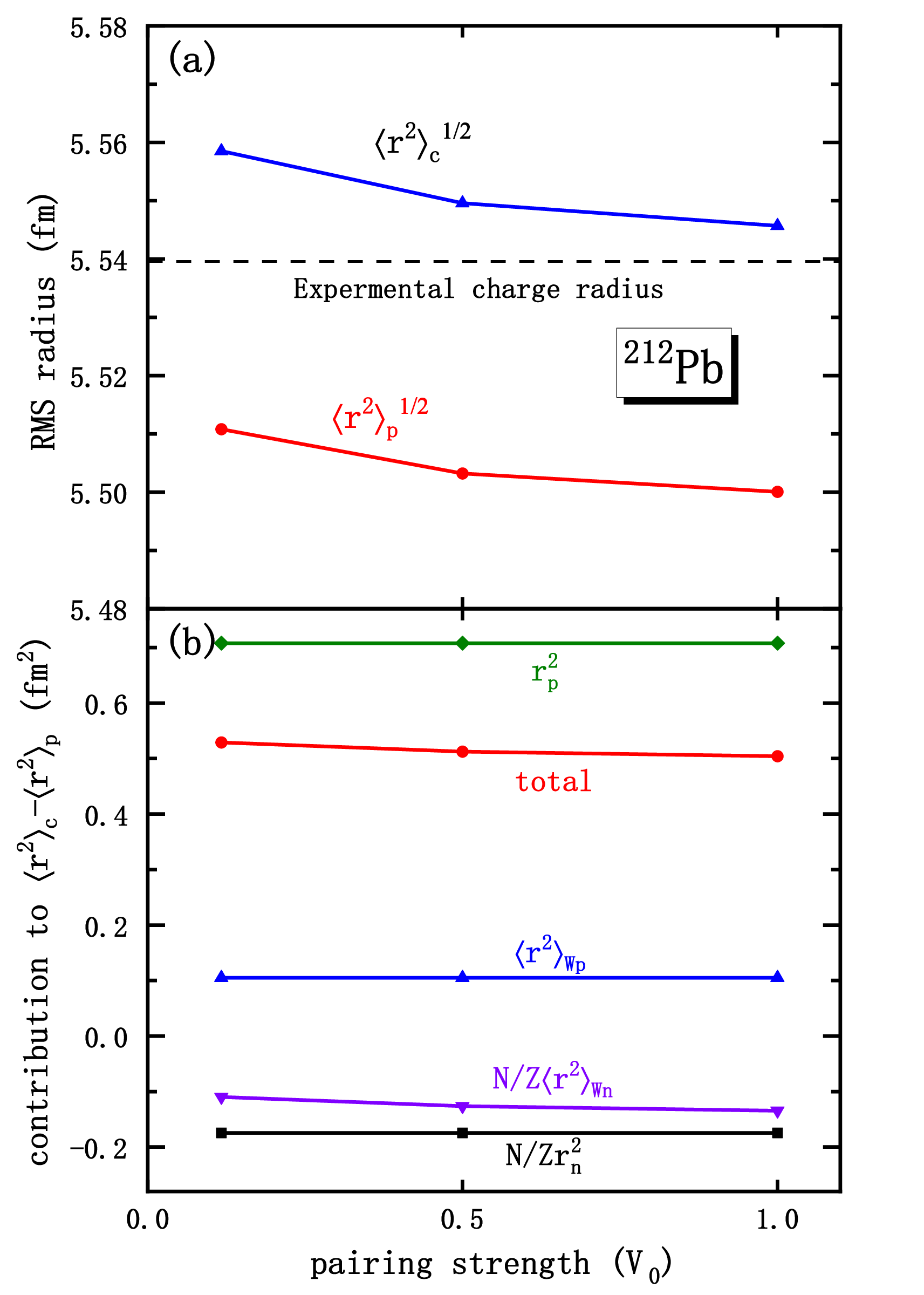}
	\caption{(Color online) (a) The charge radii and point-proton radii of $^{212}$Pb from RHB calculations \textcolor{black}{with PC-PK1 interaction} for different pairing strength. \textcolor{black}{The dashed line denotes the experimental charge radius of $^{212}$Pb.} (b) The total EM structure corrections, i.e., $\langle r^2\rangle_{ c}-\langle r^2\rangle_{ p}$, and the corresponding four components including the intrinsic nucleon contribution and the spin-orbit contributions, for $^{212}$Pb from RHB calculations with different pairing strength. }
	\label{fig:fig2}
\end{figure}

In the current study, the nuclear charge density has been constructed self-consistently from the RHB calculations with PC-PK1 interaction. 
To investigate the influence of pairing correlations on nuclear charge density and charge radius \textcolor{black}{in} open-shell nuclei, the nucleus \textcolor{black}{$^{212}$Pb} \textcolor{black}{is taken} as an example. Comparisons of charge densities and radii with different pairing strengths \textcolor{black}{are presented} in Figs.~\ref{fig:fig1} and \ref{fig:fig2}, respectively. 
As shown in Fig.~\ref{fig:fig1},  nuclear charge densities for $^{212}$Pb \textcolor{black}{with and without pairing correlations} \textcolor{black}{exhibit} \textcolor{black}{slight} differences. \textcolor{black}{Specifically, charge densities with pairing \textcolor{black}{correlations} are larger than \textcolor{black}{those} without pairing \textcolor{black}{correlations} around $4$ fm. The inset \textcolor{black}{reveals} that charge densities with pairing \textcolor{black}{correlations} \textcolor{black}{are lower} than \textcolor{black}{those} without pairing \textcolor{black}{correlations in} the surface region \textcolor{black}{($r>5.44$ fm)}.} 

In Fig.~\ref{fig:fig2}, charge radii $\langle r^2 \rangle_c^{1/2}$, point-proton radii $\langle r^2 \rangle_p^{1/2}$, and \textcolor{black}{the} four terms \textcolor{black}{of the} intrinsic electromagnetic structure corrections to mean-square charge radii in Eq.~(\ref{eq-rc}) with and without pairing correlations for $^{212}$Pb \textcolor{black}{are} given. 
It can be seen that the RHB calculations with pairing \textcolor{black}{correlations} produce smaller $\langle r^2 \rangle_c^{1/2}$ (closer to the experimental charge radius) and $\langle r^2 \rangle_p^{1/2}$ in comparison with the results without pairing \textcolor{black}{correlations}.
\textcolor{black}{As indicated by the charge densities }
in Fig.~\ref{fig:fig1} suggest that the decrease of charge radii after considering the pairing correlations is mainly related to the behavior of charge density at the surface region \textcolor{black}{($r>5.44$ fm) for $^{212}$Pb. It could be understood that charge densities obtained with pairing correlations become larger in the inner region and smaller at the surface region around 6 fm compared to the results without pairing correlations, which finally leads to smaller charge radii. It should also be noted that pairing correlation effects on the point-proton and charge radii can vary for different nuclei.}
One can see that the intrinsic proton and neutron contributions are constant, and the nucleon spin-orbit contributions are slightly changed. The total intrinsic EM structure corrections, i.e., $\langle r^2\rangle_c-\langle r^2\rangle_p$, decrease somewhat with increasing pairing strength. Thus, the changes of charge radii versus pairing strength for $^{212}$Pb are mainly originated from the charges of point-proton radii $\langle r^2\rangle_{ p}^{1/2}$.

\begin{figure*}
	\centering
	\includegraphics[width=17cm]{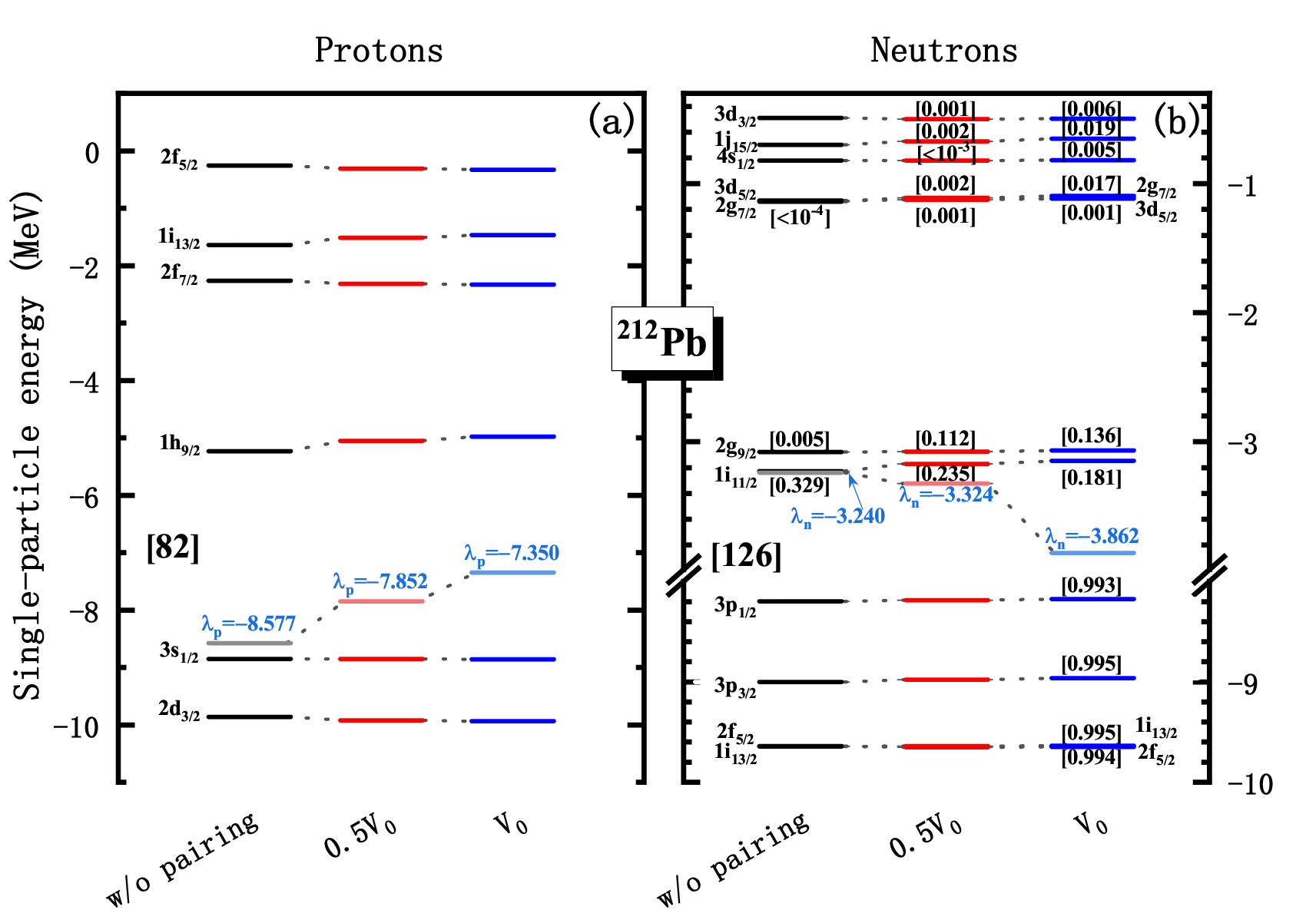}
	\caption{(Color online) The neutron and proton single-particle levels \textcolor{black}{around fermi surface} (occupation probability are labeled in square brackets) for $^{212}$Pb with pairing strengths 0.5$V_0$ and $V_0$ in canonical basis from the RHB calculations \textcolor{black}{with PC-PK1} interaction, in comparison with the calculations without pairing \textcolor{black}{correlations}. \textcolor{black}{The neutron and proton Fermi surface ($\lambda_n$ and $\lambda_p$) are also shown for comparison. See the text for details}. 
 }
	\label{fig:fig3}
\end{figure*}
 
 As pointed out in Ref.~\cite{PhysRevC.104.064313}, pairing correlations modify the occupation of different subshells. 
 To make further investigation, the neutron and proton single-particle levels and their occupation probabilities for $^{212}$Pb with pairing strengths 0.5$V_0$ and $V_0$ as well as without pairing correlations are compared in Fig.~\ref{fig:fig3}. 
 The single-particle levels \textcolor{red}{are obtained as those of }
 the canonical basis~\cite{Meng1998NPA}. 
 It should be noted that the results without pairing correlations are obtained by taking a quite small pairing strength \textcolor{black}{for both proton and neutron}, while it still produces a weak \textcolor{black}{neutron} pairing energy since four valence neutrons cannot fully occupy the orbital $1i_{11/2}$ which is \textcolor{black}{the} closest to the Fermi surface. 
 In that case, the occupation probabilities of neutron orbitals $1i_{11/2}$ and $2g_{9/2}$ are $0.329$ and $0.005$, respectively, and the occupation probabilities of the other neutron orbitals above the Fermi surface are less than $10^{-4}$. The occupation probability of the respective orbital is defined in such a way that it is equal to 1 or 0 when a given orbital is either fully occupied or empty.

 As pairing correlations are included with pairing strengths 0.5$V_0$ and $V_0$, the occupation probabilities of neutron orbitals around the Fermi surface are changed: increase for orbitals above the Fermi surface and decrease for orbitals below the Fermi surface, where the neutron orbital $1i_{11/2}$ is an exception.  In addition, single-neutron levels and the corresponding Fermi surface are also changed. \textcolor{black}{Note that in the present RHB calculations, the neutron orbital $2g_{9/2}$ lies higher than $1i_{11/2}$ in $^{212}$Pb, different from the Hartree-Fock-Bogoliubov calculations in Ref.~\cite{PhysRevC.91.021302}. It has been also discussed that the occupation probabilities on $2g_{9/2}$ and $1i_{11/2}$ and the degeneracy between these two levels are important for reproducing the odd-even staggering in the isotope shift of the Pb nuclei at $N = 126$~\cite{PhysRevC.91.021302}. Considering the $2g_{9/2}$ and $1i_{11/2}$ levels are pseudospin partners referred to pseudospin symmetry~\cite{SHANG2024138527}, manifested as a relativistic symmetry of the Dirac Hamiltonian, it is interesting to study the evolution of these two levels and the odd-even staggering of charge radii in Pb isotopic chain within relativistic density functional theory in the future.}
 Although the single-proton energies have been changed with increasing pairing strength, the $Z$ = 82 magicity is kept, and the occupations of proton orbitals \textcolor{black}{are} \textcolor{black}{unchanged, i.e., the proton orbitals below the Fermi surface $\lambda_p$ are fully occupied and the orbitals above the Fermi surface are empty.} 
\textcolor{black}{It should be noted that} the modified occupation of neutron orbitals would influence proton single-particle wave functions and thus change the point-proton RMS radii as shown in Fig.~\ref{fig:fig2}.
 
\begin{figure*}
 	\centering
 	\includegraphics[width=14cm]{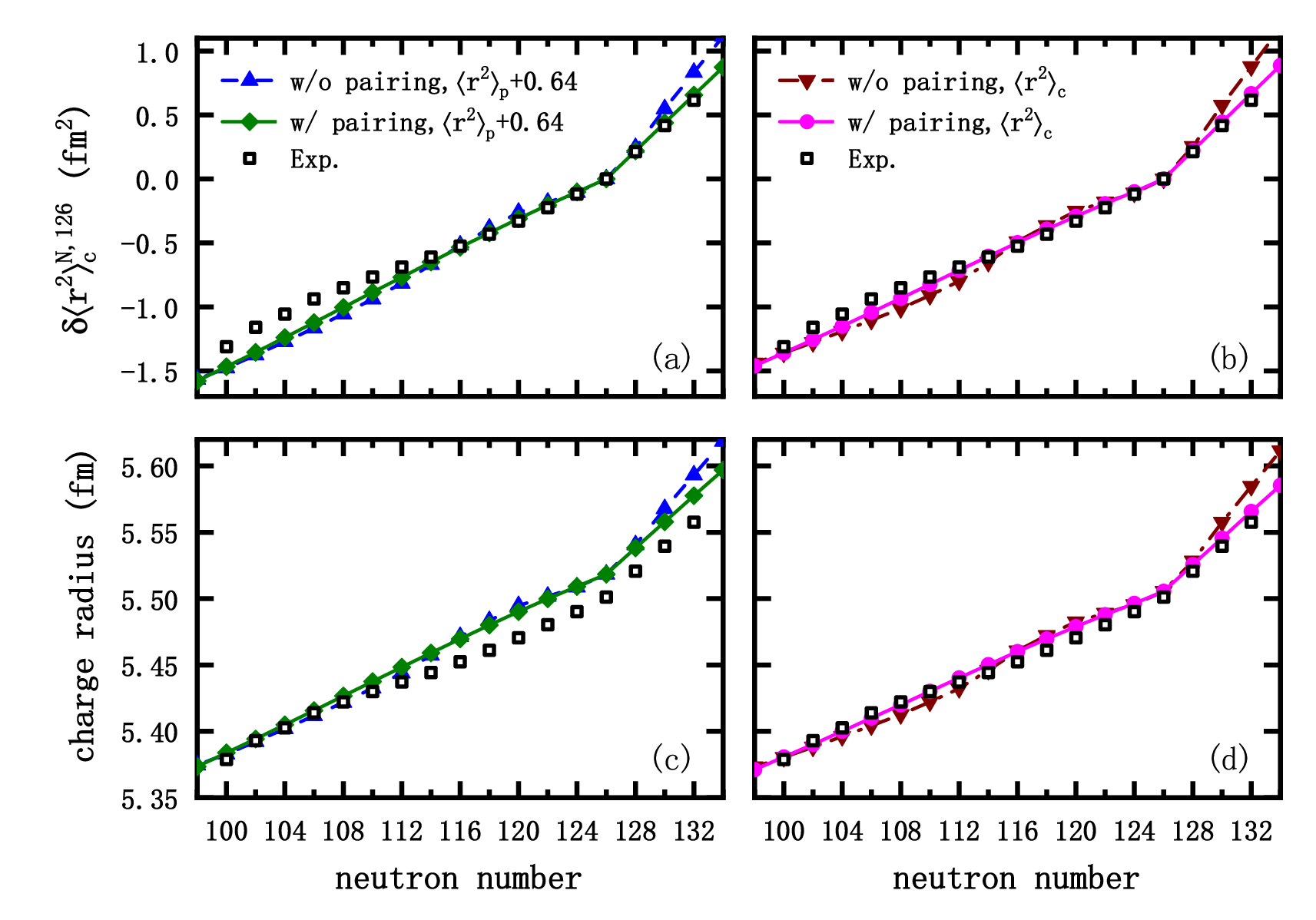}
 	\caption{(Color online) \textcolor{black}{The differential mean-square charge radius $\delta\langle r^2\rangle_{\rm c}^{N,126}$ and charge radii of even-even Pb isotopes obtained in the RHB calculations \textcolor{black}{with PC-PK1} with and without pairing correlations, in comparison with corresponding experimental data~\cite{ANGELI201369}. The calculated results with only intrinsic proton contribution (left panels) and with the intrinsic EM structure corrections (right panels) are also given for comparison.}
 	}
 	\label{fig:fig4}
\end{figure*}

To \textcolor{black}{further} study the contribution from intrinsic EM structure corrections and pairing correlations, the charge radii of even-even Pb isotopes obtained from the RHB calculations with PC-PK1 interaction are given in Fig.~\ref{fig:fig4}. In addition, the differential mean-square charge radius, defined as the difference of the mean-square charge radii, $\delta\langle r^2\rangle_{\rm c}^{N,N'}=\langle r^2\rangle_{\rm c}[N]-\langle r^2\rangle_{\rm c}[N']$, is used to facilitate the quantitative comparison for the isotopic trend of charge radii, and is also presented in Fig.~\ref{fig:fig4}. In Figs.~\ref{fig:fig4}(a) and~\ref{fig:fig4}(c), the RHB results using Eq.~(\ref{eq-rc-p}) take into account \textcolor{black}{only the} intrinsic proton contribution. It is \textcolor{black}{evident} that RHB results with only intrinsic proton contribution can well reproduce the charge radius for neutron number $N\le112$ \textcolor{black}{but} overestimate the charge radius of nuclei around the neutron magic number $N=126$ about $0.01\sim0.04$ fm, \textcolor{black}{regardless of whether} the pairing correlations is considered. Moreover, the kink of differential mean-square charge radius $\delta\langle r^2\rangle_{\rm c}^{N,126}$ at $N=126$ \textcolor{black}{is} well described, while the $\delta\langle r^2\rangle_{\rm c}^{N,126}$ are underestimated for neutron number $N\le112$.

After the intrinsic EM structure corrections are considered, both charge radii and differential mean-square charge radii $\delta\langle r^2\rangle_{\rm c}^{N,126}$ are reproduced very well as shown in Figs.~\ref{fig:fig4}(b) and~\ref{fig:fig4}(d). Specifically, the RMS deviations of charge radii in Figs.~\ref{fig:fig4}(c) and~\ref{fig:fig4}(d) between RHB calculations and experimental data are 0.0175 (w/o pairing, $\langle r^2\rangle_p+0.64$), 0.0145 (w/ pairing, $\langle r^2\rangle_p+0.64$)), 0.0106 (w/o pairing, $\langle r^2\rangle_c$) and 0.0055 fm (w/ pairing, $\langle r^2\rangle_c$). 
One can also see from Figs.~\ref{fig:fig4}(a) and~\ref{fig:fig4}(b) that pairing correlations mainly improve the description of charge radii with $N > 126$ and are critical for reproducing the kink at $N = 126$. As pointed out in Refs.~\cite{PhysRevLett.110.032503, PhysRevLett.126.032502, PhysRevC.104.064313}, \textcolor{black}{the description of kink at $N = 126$ is largely determined by the occupation of two neutron orbitals $1i_{11/2}$ and $2g_{9/2}$ and the gap between them\textcolor{black}{, which are} affected by pairing correlations. \textcolor{black}{For instance}, Figs.~\ref{fig:fig2} and~\ref{fig:fig3} have shown the influences of pairing \textcolor{black}{correlations} on charge radii and the occupation of neutron orbitals for $^{212}$Pb, as discussed in Ref.~\cite{PhysRevC.104.064313}.}

\begin{figure}
	\centering
	\includegraphics[width=8.5cm]{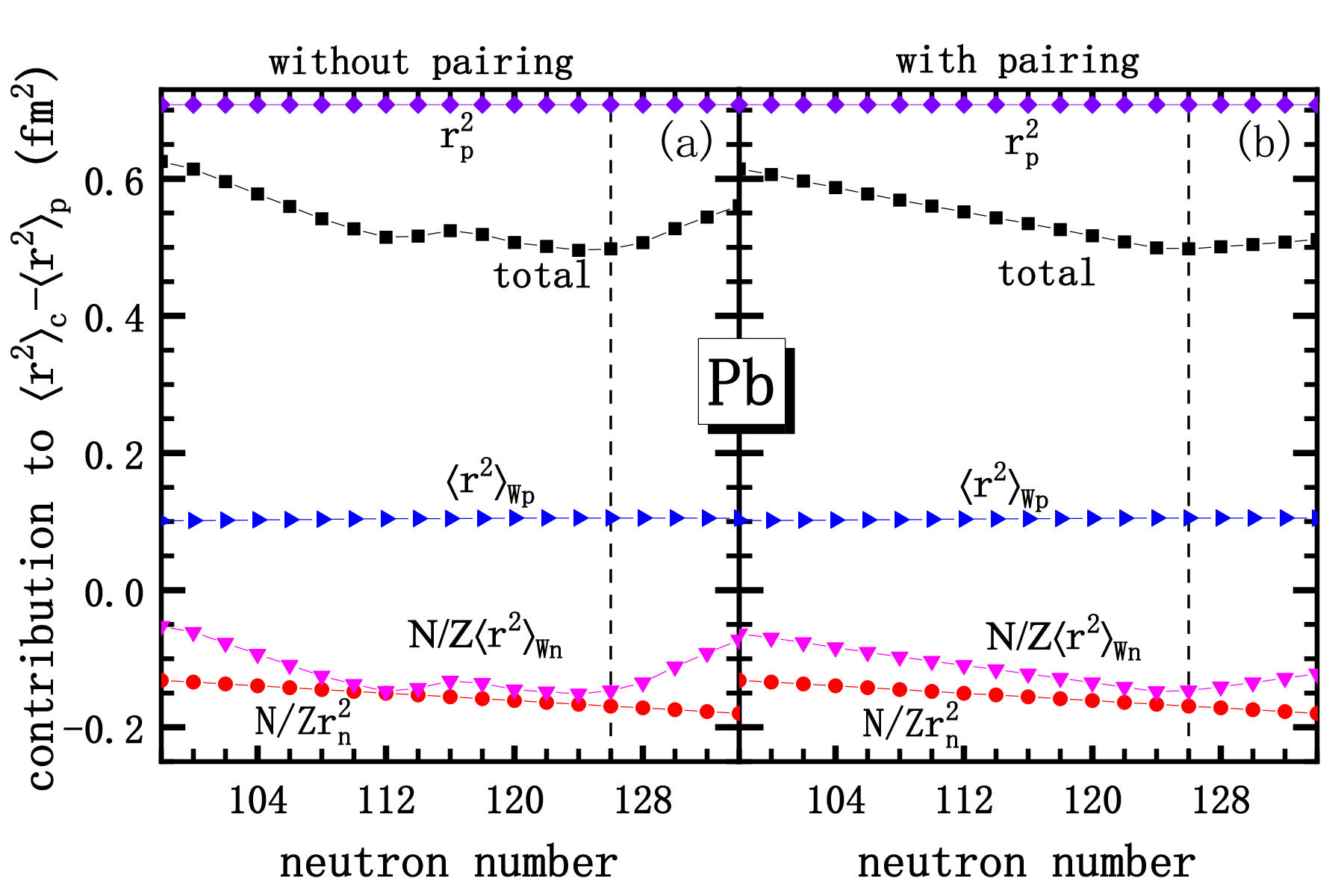}
	\caption{(Color online) The evolutions of Four corrections, i.e., the intrinsic proton (rhombus) and neutron (circle) contributions, as well as the proton (right triangle) and neutron (lower triangle) spin-orbit contributions, to $\langle r^2\rangle_{\rm c}-\langle r^2\rangle_{p}$ and the total contribution (square) along the even-even Pb isotopic chain from the RHB calculations \textcolor{black}{with PC-PK1}, without (a) and with (b) considering pairing. The black dashed lines represent $N=126$. }
	\label{fig:fig5}
\end{figure}

To further understand the influence of intrinsic EM structure, figure~\ref{fig:fig5} shows the evolution of total intrinsic EM structure corrections and \textcolor{black}{those} four corrections, i.e., the intrinsic proton and neutron contributions, as well as the proton and neutron spin-orbit contributions, with and without pairing correlations along Pb isotopic chain. \textcolor{black}{The total intrinsic EM structure corrections, i.e., $\langle r^2\rangle_{\rm c}-\langle r^2\rangle_{p}$, are between $0.5\,\mathrm{fm}^2$ and $0.6\,\mathrm{fm}^2$ in the \textcolor{black}{calculated} even-even Pb isotopes, in agreement with the non-relativistic skyrme \textcolor{black}{} calculations in Ref.~\cite{PhysRevC.103.054310}. 
The distinct difference is that $\langle r^2\rangle_{\rm c}-\langle r^2\rangle_{p}$ in the RHB calculations show a kink at $N=126$, in comparison with the decreasing trend as a function of neutron number in the skyrme calculations~\cite{PhysRevC.103.054310}.
}
It can be seen that the intrinsic proton contribution $r_{p}^2$ is constant, and the proton spin-orbit contribution $\langle r^2\rangle_{\mathrm W_{p}}$ shows a slight change. In comparison, the intrinsic neutron contribution $N/Z r^2_{n}$ decreases linearly from $-0.132$ to $-0.180$ fm$^2$ with increasing neutron number, which plays an important role in improving the predictions of the $\delta\langle r^2\rangle_{\rm c}^{N,126}$. Moreover, the neutron spin-orbit contribution $\langle r^2\rangle_{\mathrm W_{n}}$ is also significant for improving the description of $\delta\langle r^2\rangle_{\rm c}^{N,126}$. 
For example, the intrinsic neutron and the neutron spin-orbit contributions to $\delta\langle r^2\rangle_{\rm c}^{100,126}$ are $0.035$ and $0.077~\mathrm{fm}^2$, respectively, which is responsible for the prediction improved from $-1.477\mathrm{fm}^2$ (w/ pairing, $\langle r^2\rangle_p+0.64$) to $-1.359\mathrm{fm}^2$ (w/ pairing, $\langle r^2\rangle_c$), in good agreement with \textcolor{black}{the experimental data} $-1.311(13)~\mathrm{fm}^2$~\cite{ANGELI201369}, as shown in \textcolor{black}{Figs.~\ref{fig:fig4}(a) and~\ref{fig:fig4}(b)}. 

\begin{figure}
	\centering
	\includegraphics[width=8.5cm]{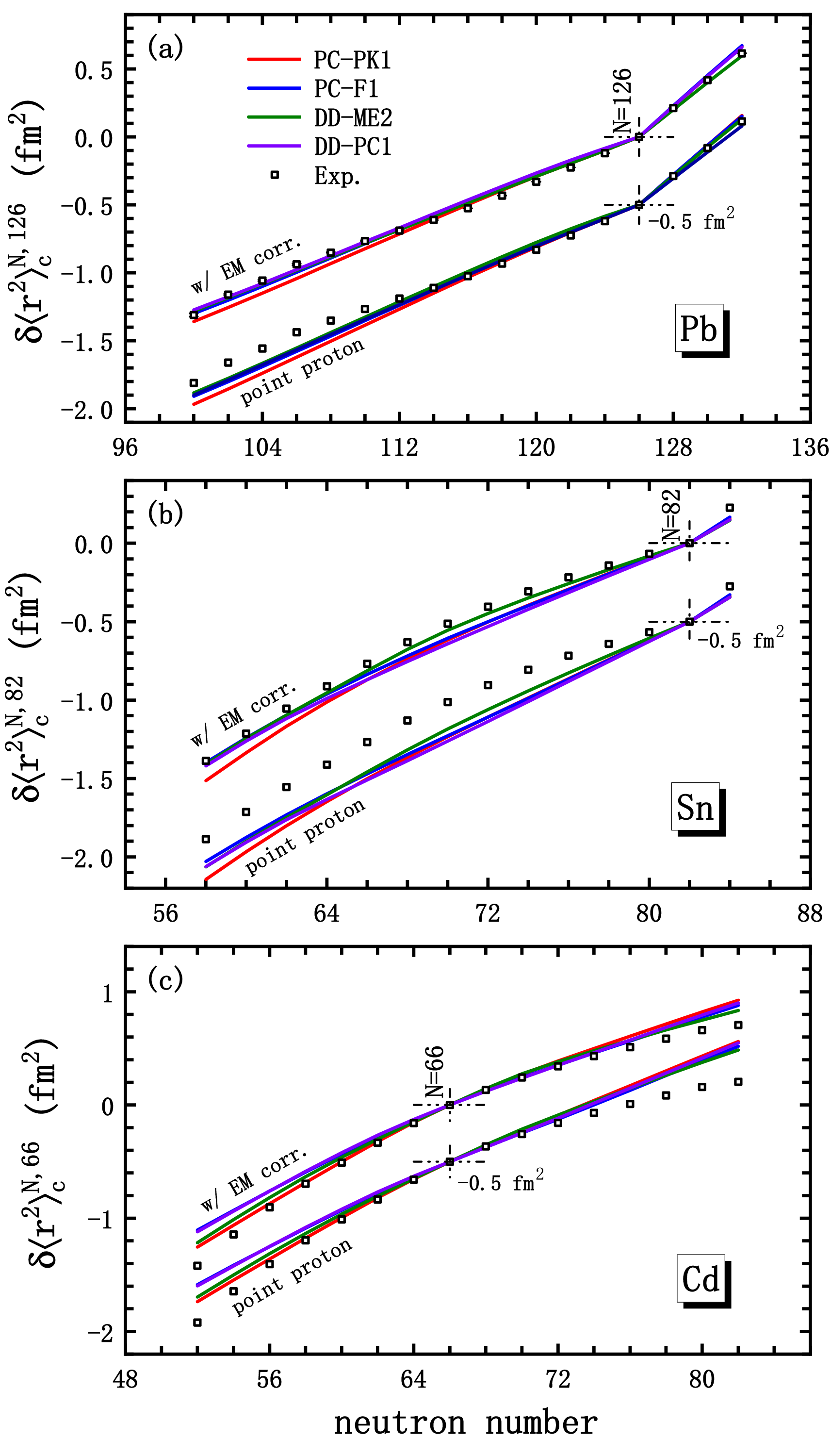}
	\caption{(Color online) \textcolor{black}{The differential mean-square charge radius $\delta \langle r^2\rangle_c^{N,N'}$ of even-even (a) Pb, (b) Sn, and (c) Cd isotopes with and without the intrinsic EM structure corrections obtained by using PC-PK1, PC-F1~\cite{PhysRevC.65.044308}, DD-ME2~\cite{PhysRevC.71.024312}, and DD-PC1~\cite{PhysRevC.78.034318} functionals, in comparison with the corresponding experimental data~\cite{ANGELI201369,PhysRevLett.122.192502,PhysRevLett.121.102501}. Note that the results of the RHB calculations without the intrinsic EM structure corrections (namely point-proton results) are shifted down by $0.5$ fm$^2$ for Pb, Sn, and Cd isotopes to compare these results on the same panel. See the text for details.}  }
	\label{fig:fig6}
\end{figure}

\begin{figure}
	\centering
	\includegraphics[width=8.5cm]{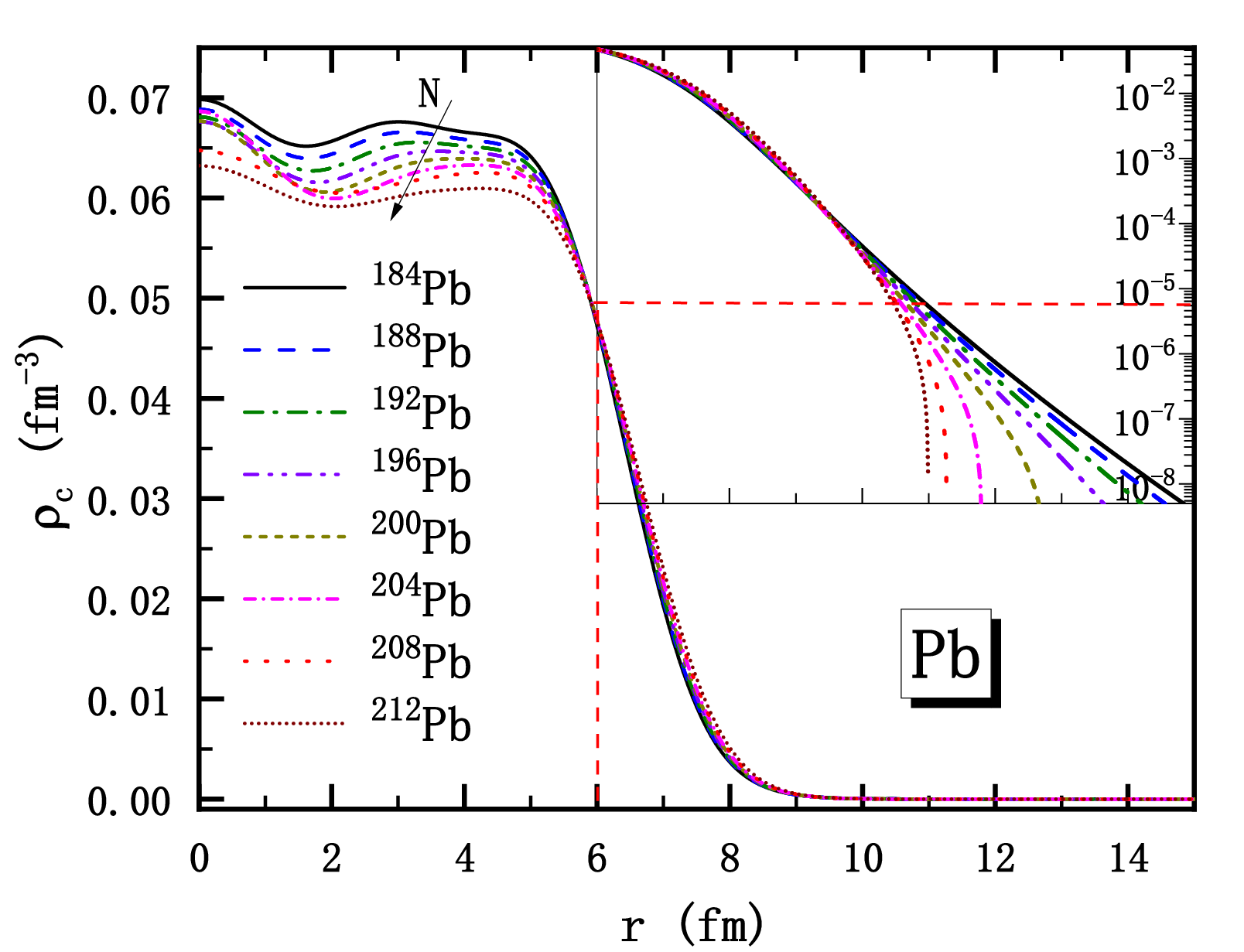}
	\caption{(Color online) Nuclear charge density distributions of Pb isotopic chain obtained from RHB calculations~\textcolor{black}{with PC-PK1}, where the thick arrows represent the evolution trend of density with neutron number. The inset shows the distributions at the surface region, i.e., 6 fm $<$ $r$ $<$ 15 fm, in a logarithmic vertical scale. }
	\label{fig:fig7}
\end{figure}

Furthermore, 
it can also be seen that the pairing correlations \textcolor{black}{slightly} affects the neutron spin-orbit correction $N/Z\langle r^2\rangle_{\mathrm W_{n}}$. 
After considering the pairing correlations, the neutron spin-orbit contribution changes more smoothly, \textcolor{black}{as pointed out in Ref.~\cite{PhysRevC.103.054310}}. 
\textcolor{black}{By comparing} the results with and without pairing \textcolor{black}{correlations} in \textcolor{black}{Figs.~\ref{fig:fig4}(a) and~\ref{fig:fig4}(b)}, it is considered that the effect of pairing correlations on the neutron spin-orbit contribution is non-negligible for the reproduction of the experimental $\delta\langle r^2\rangle_{\rm c}^{N,126}$, especially for the kink at $N=126$. %
In detail, the slope of $N/Z\langle r^2\rangle_{\mathrm W_{n}}$ curve without pairing \textcolor{black}{correlations} in Fig.~\ref{fig:fig5}(a) changes at around $N =$ 126, 116, and 112, which could result from the changes of the occupation of different \textcolor{black}{single-neutron levels}, e.g., $3p_{1/2}$, $3p_{3/2}$, $2f_{5/2}$, and $1i_{13/2}$. 
In the RHB calculations with pairing \textcolor{black}{correlations}, note that the occupations of different orbitals \textcolor{black}{are} redistributed smoothly, \textcolor{black}{and} the changes of  $N/Z\langle r^2\rangle_{\mathrm W_{n}}$ in the region of $100\le N\le 126$ are mainly related to the increase of the occupation probabilities of the neutron subshell $1i_{13/2}$.

\begin{table}
	\caption{
    \textcolor{black}{The root-mean-square (RMS) deviation $\sigma$ of the differential charge radii obtained from RHB calculations using four relativistic functionals -- PC-PK1, PC-F1~\cite{PhysRevC.65.044308}, DD-ME2~\cite{PhysRevC.71.024312}, and DD-PC1~\cite{PhysRevC.78.034318}. The upper values represent the RMS deviations without intrinsic EM structure corrections, while the lower values represent the deviations after incorporating intrinsic EM structure corrections. These deviations are compared to experimental data for the Pb, Sn, and Cd isotopic chains. }
 }\label{tab-1}
	\centering
	\setlength{\tabcolsep}{5mm}{
		\begin{tabular}{cccc}
			\toprule
			functionals       &   Pb   &   Sn   &   Cd   \\ \hline
			\multirow{2}{*}{PC-PK1} & 0.1024 & 0.1923 & 0.1448 \\
			& 0.0543 & 0.0921 & 0.0945 \\ \hline
			\multirow{2}{*}{PC-F1}  & 0.0753 & 0.1607 & 0.1601 \\
			& 0.0395 & 0.0640 & 0.1228 \\ \hline
			\multirow{2}{*}{DD-ME2} & 0.0680 & 0.1484 & 0.1316 \\
			& 0.0280 & 0.0410 & 0.0838 \\ \hline
			\multirow{2}{*}{DD-PC1} & 0.0628 & 0.1873 & 0.1678 \\
			& 0.0392 & 0.0870 & 0.1251 \\ \toprule
	\end{tabular}}
\end{table}


To further illustrate the \textcolor{black}{influence} of the intrinsic EM structure corrections \textcolor{black}{on} the description of nuclear charge radii, the differential mean-square charge radii with and without the intrinsic EM structure corrections obtained from \textcolor{black}{relativistic functionals} PC-PK1, PC-F1~\cite{PhysRevC.65.044308}, DD-ME2~\cite{PhysRevC.71.024312}, and DD-PC1~\cite{PhysRevC.78.034318} for even-even Pb, Sn, and Cd isotopes are given and compared with the corresponding experimental data in Fig.~\ref{fig:fig6}. 
In the PC-F1 calculations, the density-dependent zero-range force given by Eq.~(\ref{pair}) with \textcolor{black}{a} pairing strength $V_0=-342.5$~MeV fm$^3$ \textcolor{black}{is employed} for both neutron and proton. The DD-ME2 and DD-PC1 calculations are \textcolor{black}{performed by} the DIRHB package~\cite{NIKSIC20141808}.
It can be visibly seen that the descriptions of \textcolor{black}{the differential mean-square charge radius} for Pb, Sn, and Cd isotopic chains are all improved in these functionals \textcolor{black}{after} incorporating the intrinsic EM structure corrections. 
To exhibit the quantitative improvement, the corresponding RMS deviation \textcolor{black}{of the differential mean-square charge radius} $\sigma=\sqrt{\sum_i^n\left(\delta\langle r^2\rangle_i^{\rm Cal.}-\delta\langle r^2\rangle_i^{\rm Exp.}\right)^2/n}$ are shown in Table~\ref{tab-1}. It can be seen that the results from four relativistic functionals after considering the intrinsic EM structure corrections lead to a smaller RMS deviation $\sigma$ in comparison with the results without such corrections for Pb, Sn, and Cd isotopes.


In Fig.~\ref{fig:fig7}, the nuclear charge densities for $^{184-212}$Pb with the interval of neutron number $\Delta N=4$ \textcolor{black}{in RHB calculations with PC-PK1} are shown. One can see that the internal density decreases with increasing neutron number. It should be noted that an obvious change of charge density at $^{208}$Pb is displayed in Fig.~\ref{fig:fig7}, which is due to the shell structure effects. 
Furthermore, the density distributions at the surface region are also shown in the inset of Fig.~\ref{fig:fig7}. It is of special interest that the charge density distributions at the surface fall more rapidly with increasing neutron numbers, which are just the opposite of neutron density distributions shown in Ref.~\cite{XIA20181}. It should be noted that the intrinsic neutron contribution is responsible for the shape of nuclear charge distribution at the surface region, given negative charge distribution at the surface for a single neutron according to experimental measurement~\cite{PERDRISAT2007694}.

\section{summary and outlook}\label{sec-4}
In this paper, the contributions \textcolor{black}{of the} intrinsic EM structure, i.e., intrinsic nucleon contributions and spin-orbit density contributions, and \textcolor{black}{the effects} the pairing \textcolor{black}{correlations on} the nuclear charge density distribution and nuclear radii, are simultaneously taken into account in the RHB model. \textcolor{black}{Using} the Pb isotopes as examples, the corresponding contributions have been studied.

The charge density and charge radii of the open-shell nucleus $^{212}$Pb are influenced by the pairing correlations, as shown \textcolor{black}{with} the pairing strengths 0.5$V_0$ and $V_0$ as well as without pairing \textcolor{black}{correlations} in the RHB calculations. The total intrinsic EM structure correction \textcolor{black}{to mean-square charge radius}, i.e., $\langle r^2\rangle_{\rm c}-\langle r^2\rangle_{p}$ changes slightly with increasing pairing strength, indicating that the contributions from intrinsic EM correction are slightly influenced by pairing correlations. In \textcolor{black}{contrast}, the occupation probabilities of the single-neutron levels near the Fermi surface show visible changes with increasing pairing strength. Thus, the single-proton orbitals are changed by pairing correlations, which finally \textcolor{black}{impact} the point-proton RMS radius.

The charge radii and differential mean-square charge radii of even-even Pb isotopes obtained from the RHB calculations are compared with the corresponding experimental data. 
The results show that the corresponding RMS deviations of charge radii and kink at $N=126$ are improved after considering the intrinsic EM structure corrections and pairing correlations. 
The intrinsic neutron and neutron spin-orbit contributions are responsible for the improvement, as found by comparing the isotopic evolution of these corrections. 
The isotopic evolution of the neutron spin-orbit contribution with pairing correlations is smoother than the results without pairing correlations, because after taking into account pairing correlations, the occupations of some neutron orbits around the Fermi surface, such as $1i_{11/2}$, change smoothly with increasing neutron number. 
The influence of pairing correlations on neutron spin-orbit contribution to mean-square charge radii is non-negligible for the prediction of $\langle r^2\rangle_c^{N,126}$, especially for the kink at $N=126$. The charge density distributions for Pb isotopes in the inner and surface regions are also shown. 
The curve from $^{204}$Pb to $^{208}$Pb shows a change, implying shell structure effects. 
In addition, the intrinsic neutron contribution can be seen from the rapid fall at the surface on the neutron-rich side.

By performing RHB calculations with relativistic functionals, PC-PK1, PC-F1, DD-ME2, and DD-PC1, it is found that the descriptions of the evolution of charge radii for Pb, Sn, and Cd isotopic chains are also improved after taking into account the intrinsic EM structure corrections. This further indicates that the intrinsic EM structure corrections could make an effect for accurately describing the evolution of nuclear charge radius.

In the future, the study of charge radii for odd-$A$ and odd-odd nuclei with the intrinsic EM structure correction is also of interest, apart from the even-even nuclei. 
Moreover, the intrinsic EM structure contributions for charge radius are expected to be considered in the fitting protocol to improve the nuclear chart's global description of charge radii. In Ref.~\cite{PhysRevC.107.054307}, it is mentioned that beyond-mean-field effects are also important in describing the charge radii. Thus, the intrinsic EM structure corrections and beyond-mean-field effects are expected to be incorporated. Investigations in these directions are in progress.

\begin{acknowledgments}
This work was supported by the National Natural Science Foundation of China (Nos. 12475119 and 11675063), the Natural Science Foundation of Jilin Province (No. 20220101017JC) and the Key Laboratory of Nuclear Data Foundation (JCKY2020201C157).
\end{acknowledgments}




\bibliography{apssamp}

\end{document}